\documentclass[a4paper,11pt]{article}
\usepackage{pos}
\usepackage{subcaption}
\usepackage{graphicx}
\usepackage{amsmath}
\usepackage{float}
\usepackage{caption}

\title{Probing Instanton Dynamics in the Pion Vector Form Factor with Wilson Flow}
\author*[a]{Vaibhav Chahar}
\author[b]{Piotr Korcyl}

\affiliation[a]{Doctoral School of Exact and Natural Sciences, Jagiellonian University\\
prof. {\L}ojasiewicza 11, 30-348 Krak\'ow, Poland}

\affiliation[b]{Institute of Theoretical Physics, Jagiellonian University\\
prof. {\L}ojasiewicza 11, 30-348 Krak{\'o}w, Poland}

\emailAdd{vaibhav.chahar@doctoral.uj.edu.pl}
\emailAdd{piotr.korcyl@uj.edu.pl}

\abstract{Instanton liquid model is believed to capture the main features of vacuum QCD dynamics. Recently, multiple predictions for hadron structure functions have been derived and compared with experimental measurements and lattice QCD calculations, finding a general agreement. In order to explore the precision of the instanton liquid model, one has to compare its predictions with non-perturbative simulations in a regime dominated by instanton dynamics.
This has been performed for two gluon-sensitive observables: the gluon Green's function and the strong running coupling constant. In this contribution, we propose to study a fermionic observable, the pion electromagnetic form factor, for which instanton liquid model predictions have been discussed in Phys.Rev.D 109, 074029. %\cite{Liu:2023fpj}. 
We use the Wilson flow to single out the dominant contribution from the instantons out of a lattice QCD configuration ensemble. We describe the details of our numerical setup, and some first, preliminary results.}

\FullConference{%
 The 42th International Symposium on Lattice Field Theory, LATTICE2025\\
  2-8 November 2025, Tata Institute of Fundamental Research (TIFR), Mumbai, India
}

\begin{document}
\maketitle

\section{Introduction}

The understanding and description of the internal structure of hadrons is a challenging problem with a lot of activity on both the experimental and theoretical sides. Because of its non-perturbative character, the framework of numerical lattice Quantum Chromodynamics \cite{gattringer} is often employed to make reliable theoretical predictions. Simple pieces of information, i.e., electromagnetic radius of a hadron, can be estimated from matrix elements of appropriate current operators \cite{PhysRevD.79.074506}. An alternative, although approximate, approach is offered by the instanton liquid model \cite{RevModPhys.70.323}, which builds upon our theoretical understanding of the QCD vacuum. The calculation relies on the assumption that at low energy, the properties of hadrons result from the vacuum properties of QCD that can be modelled by a dense ensemble of classical, instanton solutions to the Yang-Mills equations of motion. Recent calculations of the electromagnetic radius of the pion using this approach have been discussed in Ref.\cite{Liu:2023fpj}. From the general theoretical perspective, it is an interesting question to compare the two approaches and thus validate our understanding of the QCD vacuum structure. In the framework of lattice QCD, the Wilson flow \cite{Luscher:2009eq,Luscher:2010iy,Narayanan_2006} provides a tool that can enhance the instanton contributions to the QCD vacuum. It has been demonstrated that for sufficiently large flow time, the ultraviolet fluctuations of gauge degrees of freedom are smeared out and the configurations are dominated by instanton-like structures \cite{Athenodorou:2016gsa,Athenodorou:2018jwu}. Hence, non-perturbative, fully controlled calculations can be made in this regime of QCD dominated by instanton dynamics. In this work, we propose to study as a test observable the pion electromagnetic form factor as a function of the Wilson flow time. As a result, ultimately, we will be in the position of reliably comparing the predictions of the instanton liquid model against lattice QCD. 
%Hence, we will make a significant step towards confirming or discarding the current physical picture of the QCD vacuum. If successful, we may gain computationally cheaper access to the hadron's structure functions. Our results may have a significant phenomenological impact in view of the Electron Ion Collider whose primary role will be to experimentally measure hadron structure functions and thus unveil the internal dynamics of hadrons.

In this contribution, we describe our initial efforts to study the pion electromagnetic form factor $F_{\pi}(q^2)$ as a function of the Wilson flow $t$. We present our preliminary estimate on a single ensemble only.
The main difficulty that we address here is the appropriate strategy to compare $F_{\pi}(q^2,t)$ at different flow times. $F_{\pi}(q^2)$ depends on the quark masses through the pion mass, $F_{\pi} = F_{\pi}(m_{\pi})$. At positive flow, the pion mass changes with the flow time $t$, i.e., $F_{\pi}(m_{\pi}) = F_{\pi}(m_{\pi}(t), t)$. We would like to disentangle these dependencies and access $F_{\pi}(\bar{m}_{\pi},t)$ for some chosen pion mass $\bar{m}_{\pi}$. This leads, in particular, to the dependence of $\kappa_{m_{\pi} = \bar{m}_{\pi}}(t)$ used in the evaluation of $F_{\pi}$, such that the resulting effective pion mass stays constant as we vary the flow time. One of our results is the determination of $\kappa_{m_{\pi} = \bar{m}_{\pi}}(t)$ in a wide range of $t$. A number of other parameters should be tuned accordingly to maintain $O(a)$-improvement. In order to avoid complications, in this preliminary study, we set $c_{SW}$ coefficient to its tree-level value $c_{SW} = 1.0$, and keep it independent of the flow time. Hence, the resulting pion mass is larger than the one determined with the $c_{SW}$ coefficient fixed non-perturbatively. We fix the vector current normalization constant $Z_V(t)$ using a condition built out of two- and three-point functions \cite{PhysRevD.79.074506}. This corresponds to a massive normalization scheme. We provide estimates of $Z_V(t)$ for up to large flow times. Our determinations of $\kappa_{\bar{m}_{\pi}}(t)$ and $Z_V(t)$ interpolate between known values at $t=0$ and tree-level predictions at large $t$, confirming that the gauge field is transformed towards a unit field. Note that this means that the value of the $c_{SW}$ coefficient becomes irrelevant at large flow times, as the clover operator vanishes on the unit field.

In the rest of this contribution, we provide details on the gauge field ensemble used in the study and describe our numerical setup for the evaluation of the pion electromagnetic form factor. Afterwards, we discuss the numerical results. We show the two-point and three-point functions at different flow times, ranging from $t=t_0$ up to $t=20 t_0$. The upper limit is dictated by Ref.~\cite{Athenodorou:2016gsa,Athenodorou:2018jwu} where the changes in the instanton density were studied with the flow time. We will use flow times up to $t = 5 t_0$, which correspond to densities expected in the instanton liquid model \cite{Liu:2023fpj}. We present the dependence of $\kappa_{\bar{m}_{\pi}}(t)$ and of the form factor $F_{\pi}(q^2,t)$. In the last section, we briefly discuss further prospects.

%Current results from lattice Quantum Chromodynamics have not discussed physical observables at positive flow times as physically relevant. Neither Wilson loops nor hadronic matrix elements have been studied as a function of the flow time. The traditional strategy consisted of exploiting the Wilson flow as a technical device and performing a perturbative matching of the obtained quantities back to the $t=0$ limit. Hence, a formulation of the line of constant physics for positive flow times, including renormalization and $O(a)$-improvement, is missing, prohibiting comparisons of expectation values at different flow times. 

\section{Description of numerical setup} 
\label{sec:description}

\subsection{Observable}

The observable that we propose to study is the pion electromagnetic form factor, $F_{\pi}(q^2)$, as a function of the Wilson flow time $t$. In the continuum theory, it is given by the hadronic matrix element with an insertion of the axial-vector electromagnetic current operator,
\begin{equation}
    \label{eq. form factor}
    \langle \pi^+(p') | \hat{V}_{\mu}(0)|\pi^+(p)\rangle = F_{\pi}(q^2) (p + p')_{\mu},
\end{equation}
where $p$ and $p'$ are the initial and final pion four-momenta, $q^2 = (p-p')^2$ is the squared four-momentum transfer and $\hat{V}_{\mu}$ is a conserved electromagnetic current operator on the lattice,
\begin{equation}
    \hat{V}_{\mu}(x) = Z_V \frac{1}{2} \Big[ \bar{u}(x) \gamma_{\mu} u(x) - \bar{d}(x) \gamma_{\mu} d(x) \Big],
\end{equation}
$Z_V$ is the renormalization constant of the isovector part of the vector current. The matrix element Eq.~\eqref{eq. form factor} can be accessed on the lattice from the two- and three-point correlation functions \cite{PhysRevD.79.074506},
\begin{align}
C^{\pi}(\tau, \vec{p})
&=
\sum_{x,z}
\langle O_{\pi}(x)\, O^{\dagger}_{\pi}(z) \rangle\,
\delta_{\tau,\tau_x-\tau_z}\,
e^{-i\vec{p}\cdot(\vec{x}-\vec{z})},
\label{eq:Cpi}
\\
C^{\pi\pi}_0(\tau, \tau', \vec{p}, \vec{p}\,')
&=
\sum_{x,y,z}
\langle O_{\pi}(y)\, V_0(x)\, O^{\dagger}_{\pi}(z) \rangle\,
\delta_{\tau,\tau_x-\tau_z}\,
\delta_{\tau',\tau_y-\tau_z}\,
e^{-i\vec{p}\cdot(\vec{x}-\vec{z}) - i\vec{p}\,'\cdot(\vec{x}-\vec{y})}.
\label{eq:Cpipi0}
\end{align}
where $V_0(x) = \bar{u}(x) \gamma_0 u(x)$ and $O^{\dagger}_{\pi}(z) = \bar{u}(z) \gamma_5 d(z)$ is the interpolating operator for the $\pi^+$. 
We follow Ref.~\cite{Capitani1999} and build ratios of two-point and three-point functions to cancel out  common normalizations and to reduce the statistical noise, 
\begin{equation}
R(\tau,\tau',\vec{p}, \vec{p}\,') =
\frac{C^{\pi\pi}_0(\tau,\tau', \vec{p}, \vec{p}\,')}{C^{\pi}(\tau', \vec{p}')}
\sqrt{
\frac{
C^{\pi}(\tau'-\tau,\vec{p})\,
C^{\pi}(\tau,\vec{p}\,')\,
C^{\pi}(\tau',\vec{p}\,')
}{
C^{\pi}(\tau'-\tau,\vec{p}\,')\,
C^{\pi}(\tau,\vec{p})\,
C^{\pi}(\tau',\vec{p})
}
}.
\end{equation}
The source is set at timeslice 0, and the final pion state is localized on the $\frac{1}{2}T$ timeslice. The ratio is multiplied by a factor 2 when extracting pion matrix elements. We keep the final pion state at rest. The vector current insertion is at $\tau$.
% Note that the time $\tau$ is the time component of the four-vector $x$ and has nothing to do with the gradient flow time $t$. 
Using the lattice normalization of pion states and the zeroth component of the vector current, we have
%It can be shown that \cite{PhysRevD.79.074506}
\begin{equation}
    \label{eq. pion form factor}
    2\sqrt{E_{\vec{p}}\,E_{\vec{p}'}}\;Z_V\left\langle \pi(\vec{p}') \left| V_0(0) \right| \pi(\vec{p}) \right\rangle
= 2\bar{p}_{0}\, F_{\pi}(q^2), 
\end{equation}
with $\bar{p}^{\mu{}} = \frac{1}{2} (p^{\mu}\,' + p^{\mu{}})$ and hence
\begin{equation}
    \langle \pi(\vec{p}\,') | V_0(0) | \pi(\vec{p}) \rangle
= 2  R({\tau},\tau',\vec{p}, \vec{p}\,').
\end{equation}
The vector current normalization factor is defined by \cite{PhysRevD.79.074506},
\begin{equation}
    \label{eq. ZV}
    Z_V = \frac{C^{\pi}_0(\tau', 0)}{C^{\pi \pi}_0(\tau, \tau', 0)}.
\end{equation}
This defines a massive renormalization scheme and therefore $Z_V \ne 1$ on the unit field. Rather, one should interpret it as $Z_V = Z_V^{\textrm{massless}} \big( 1 + Z_V^{\textrm{massive}}(m_{\pi}) \big)$, where $Z_V^{\textrm{massless}} = 1$ on the unit field and $Z_V^{\textrm{massive}}(0) = 0$.

%\paragraph{Flowed pion electromagnetic form factor and the continuum limit\\}

% Once the pion form factor Eq.~\eqref{eq. pion form factor} is evaluated on a set of gauge field ensembles with varying lattice spacing, pion mass, and volume, one can perform a combined extrapolation to the continuum limit, physical pion mass, and infinite volume, thus obtaining the phenomenologically relevant estimate. The same procedure can be applied to a set of gauge field ensembles flowed with the Wilson flow. .

% The line of constant physics will be defined using appropriate Ward-Takahashi identities which provide access to renormalized operators protected by global symmetries \cite{Vladikas:2011bp}. The $\mathcal{O}(a)$-improvement may require additional simulations to be performed at vanishing quark mass which can be done using the Schroedinger Functional formulation of lattice QCD \cite{Sint_1995,L_scher_2006,Heitger:2020mkp}. \\

\subsection{Gauge field ensemble}

For the calculation, we use the publicly available PACS-SC ensemble of 200 gauge configurations \cite{Aoki:2008sm}. The QCD action is discretized on a hyper-cubic lattice with lattice spacing $a = 0.0907$ fm and size $32^3$$\times 64$. The gauge part of the action is the Iwasaki gauge action, and the fermionic part of the action is the non-perturbatively O(a)-improved Wilson action with $N_f = 2 + 1$ dynamical quarks. The pion mass used is $m_{\pi} = 409.7(7)$ MeV. Further details of this ensemble are summarized in Tab.~\ref{tab. ensemble}.\\

\begin{table}[H]
\centering
\begin{tabular}{ccccccccccccc}
\hline
 & $\beta$ & $\kappa_l$ & $L/a$ & $T/a$ & $c_{\mathrm{sw}}$ & $N_G$ &  $a$ [fm] & $m_\pi$ [MeV]  & $Z_V$ \\
\hline
& 1.90 & 0.13754 & 32 & 64 & 1.715 & 200 & 0.0907(13) & 409.7(7) & 0.7354(37) \\
\hline
\end{tabular}
\caption{Combined bare and derived parameters for the ensemble used in this study as in Ref.~\cite{Aoki:2008sm}. \label{tab. ensemble}}
\end{table}

\section{Preliminary results}

% The correlator in Eq.~\eqref{eq:Cpi} written as after Wick contractions:
% \begin{equation}
% C^{\pi}(\tau,\vec{p})
% =
% - \sum_{x,z} e^{-i\vec{p}\cdot(\vec{x} - \vec{z})}\,
% \mathrm{Tr}\!\left[
% D_u^{-1}(x)\,\gamma_1\Gamma^{\dagger}_{4}\gamma_5\,
% D_d^{-1}(z)\,\gamma_5
% \right] ,
% \label{eq:C2pi}
% \end{equation}
% Dirac propagator (matrix notation) in Eq. (20) for a quark is defined by Eq. (21)
% \begin{equation}
% D = 1 - \kappa H,
% \qquad
% \kappa = \frac{1}{2\,(am_q+4)} \,.
% \label{eq:DiracOp}
% \end{equation}
% where H is a hopping matrix and $\kappa$ is hopping parameter.
% \begin{equation}
% H(n|m)_{\alpha\beta}
% =
% \sum_{a,b}
% \sum_{\mu=\pm 1}^{\pm 4}
% \left(1 - \gamma_\mu\right)_{\alpha\beta}\,
% U_\mu(n)_{ab}\,
% \delta_{n+\hat{\mu},\,m} \,.
% \label{eq:Hopping}
% \end{equation}

\begin{figure}
    \centering
    \includegraphics[width=0.7\textwidth]{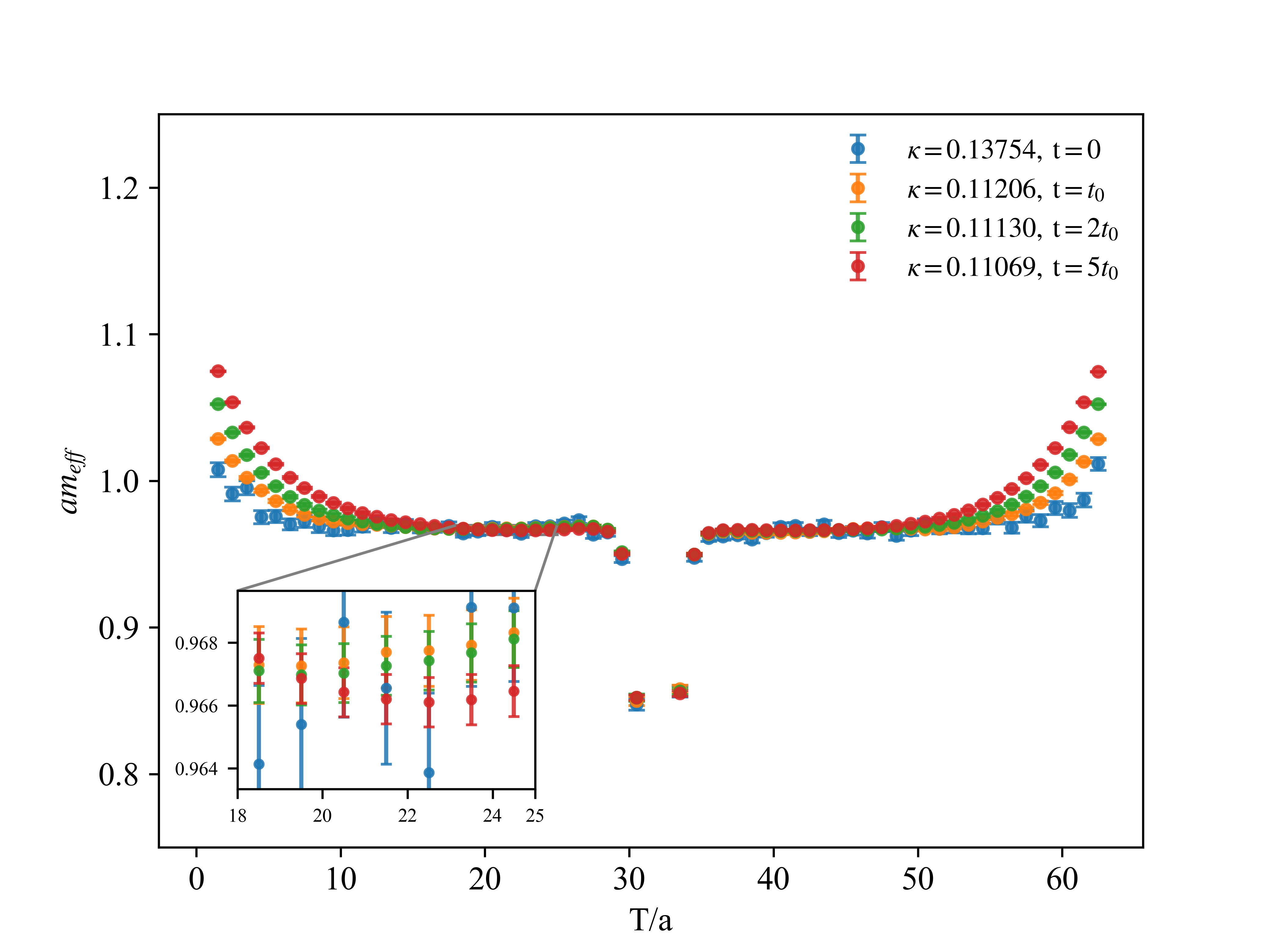}
    \caption{The pion effective mass at different Wilson flow time $t$ after tuning the $\kappa$ parameter. Although the structure of excited states is different, the ground state energies are matched within the statistical uncertainties. The range of $t$ corresponds roughly to the range investigated in Ref.~\cite{Athenodorou:2016gsa}. }
    \label{fig:kappa_eff}
\end{figure}

\begin{figure}
    \centering
    \includegraphics[width=0.65\textwidth]{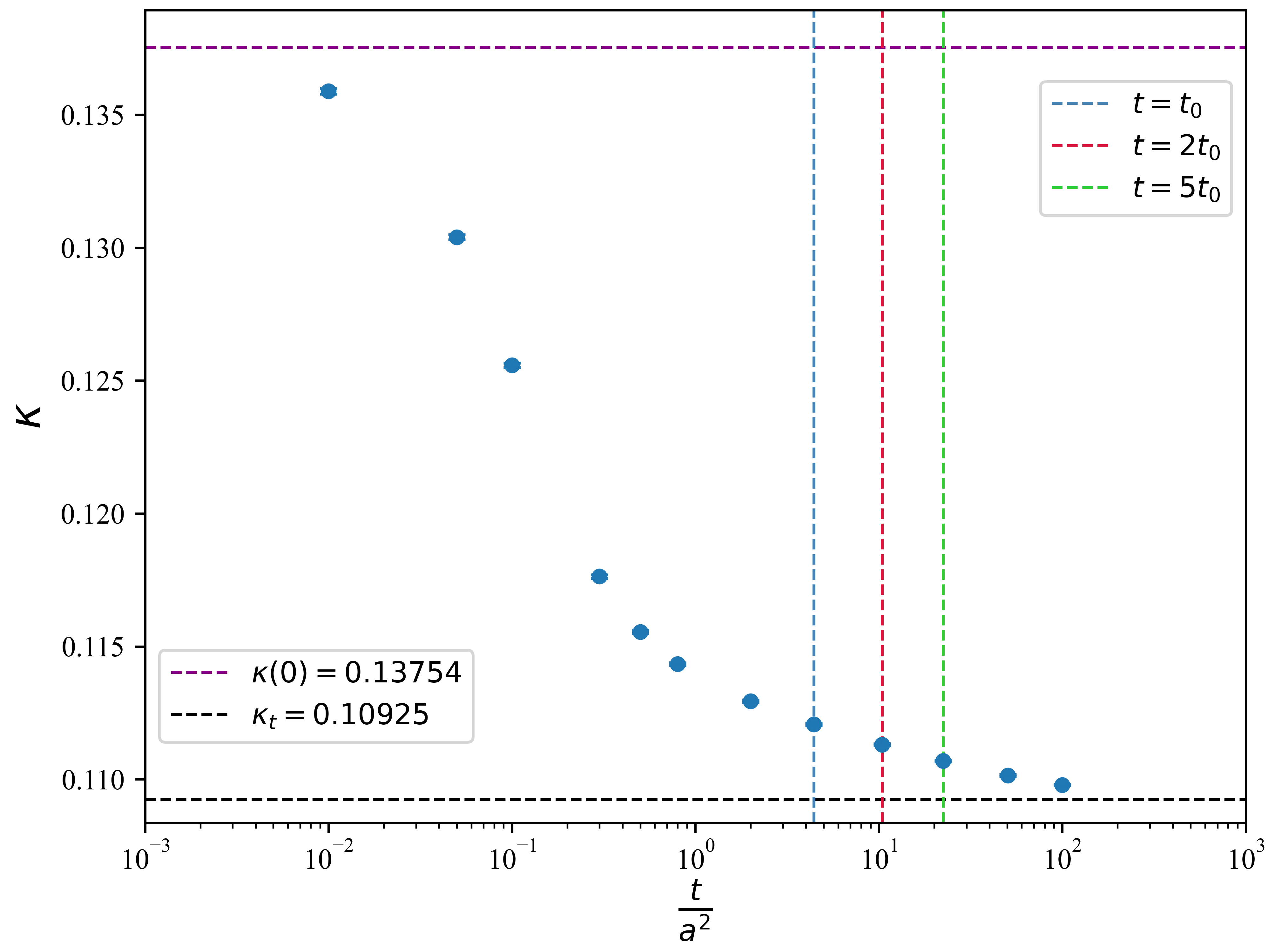}
    \caption{The value of the $\kappa$ parameter $\kappa^{am_{\pi}=0.9668}(t)$, for which the pion mass in lattice units remains constant, as a function of the Wilson flow time $t$. The violet, horizontal dashed line denotes the value of $\kappa$ of the sea quarks in the ensemble, whereas the black, horizontal dotted line represents the tree-level value estimation. Vertical lines denote specific values of the Wilson flow time $t$ for which the pion electromagnetic form factor was estimated as shown in Fig. \ref{fig:form_factor}. It shows that the measurements of the form factor were performed in the regime where $\kappa$ was already close to its tree-level value. The statistical uncertainties are smaller than the symbol size.}
    \label{fig:kappa_corr_flow}
\end{figure}

The preliminary results presented in this contribution consist of several steps. First, we evaluated the pion two-point correlation function for various values of $\kappa$ and Wilson flow time $t$. To each correlation function, we performed a two-state fit of the form 
\begin{equation}
C^{\pi}(\tau, 0)
=
A_0^{\,2}\,e^{-E_0 \tau}
+
A_1^{\,2}\,e^{-E_1 \tau}
+
A_0^{\,2}\,e^{-E_0 (T-\tau)}
+
A_1^{\,2}\,e^{-E_1 (T-\tau)} \,.
\label{eq:C2pt_twostate}
\end{equation}
to extract the pion ground state energy. By interpolating between successive values of $\kappa$ for a given Wilson flow time $t$, we estimated the functional dependence of $\kappa(t)$ on $t$ such that the resulting pion mass remains constant. We demonstrate that in Fig. \ref{fig:kappa_eff}, where we plot the effective masses defined as
% \begin{equation}
% a\,m_{\mathrm{eff}}\!\left(\frac{\tau}{a}+\frac{1}{2}\right)
% =
% \log\!\left[
% \frac{C^{\pi}(\tau/a,0)}{C^{\pi}(\tau/a+1,0)}
% \right] .
% \label{eq:meff}
% \end{equation}
\begin{equation}
a\,m_{\mathrm{eff}}\!\left(\tau+\frac{1}{2}\right)
=
\log\!\left[
\frac{C^{\pi}(\tau,0)}{C^{\pi}(\tau+1,0)}
\right] .
\label{eq:meff}
\end{equation}
at four different values of the Wilson flow time. We notice that with increasing $t$, the contribution of excited states increases, which may be a consequence of the quark interaction becoming weaker. After repeating this exercise for a wide range of $t$ we obtained the function $\kappa(t)$ shown in Fig. \ref{fig:kappa_corr_flow}. We observe that $\kappa(t)$ changes dynamically for Wilson flow time $t/a^2$ around $0.1$ from the sea quark value towards the tree-level value. As expected from the properties of the Wilson flow, the ultraviolet fluctuations of the gauge field are smeared out, and the resulting configuration increasingly resembles the unit field configuration.  
Knowing $\kappa(t)$, we proceed with the evaluation of the form factor using Eq.~\eqref{eq. pion form factor}. The estimated normalization factor $Z_V(t)$ as defined in Eq.~\eqref{eq. ZV} shows a dynamics similar to $\kappa(t)$. It changes dynamically at small Wilson flow time $t$ and tends to its tree-level value afterwards. 

\begin{figure}
    \centering
    \includegraphics[width=0.7\textwidth]{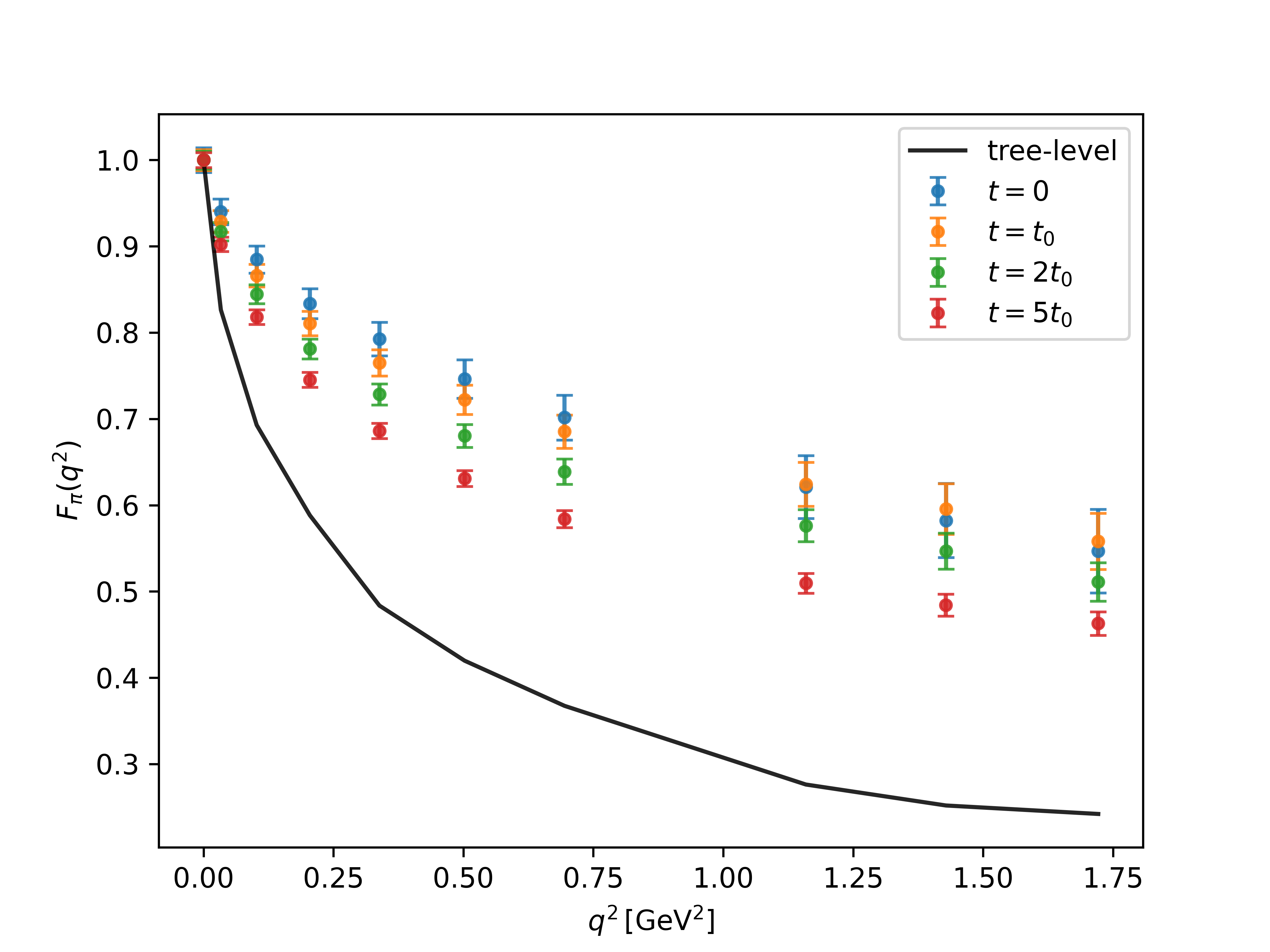}
    \caption{The pion electromagnetic form factor $F_{\pi}(q^2,t)$ as a function of the momentum transfer $q^2$ in GeV for different values of the Wilson flow $t$. The solid black line corresponds to the tree-level estimation obtained by calculating the two- and three-point functions on a unit field. }
    \label{fig:form_factor}
\end{figure}

The final results for the pion electromagnetic form factor are shown in Fig.~\ref{fig:form_factor}. We plot the data without the flow, i.e., at $t=0$ and at the three chosen values of $t$. The black solid line denotes the tree-level expectation. We notice that the form factor shows different dynamics in terms of the dependence on $t$. The values of $F_{\pi}(q^2,t>0)$ remain relatively close to $F_{\pi}(q^2,t=0)$ and tend slowly towards the tree-level limit. It is tempting to interpret that result by saying that the Wilson flow quickly averages the ultraviolet fluctuations, hence $\kappa(t)$ and $Z_V(t)$ quickly reach their tree-level values. However, the form factor is more sensitive to the vacuum structure, which evolves much more slowly with the Wilson flow time. One way of confirming this hypothesis would be to calculate the instanton density evolution with $t$ on this specific ensemble of gauge configurations and contrast that with the dynamics shown by the quantities discussed above.

% \begin{figure}[htbp]
% \centering
% \begin{subfigure}[t]{0.32\textwidth}
%   \centering
%   \includegraphics[width=\linewidth]{corr_flow_4.png}
%   \caption{Pion correlator at flow time $t=t_0$}
%   \label{fig:sub1}
% \end{subfigure}\hfill
% \begin{subfigure}[t]{0.32\textwidth}
%   \centering
%   \includegraphics[width=\linewidth]{corr_flow_10.png}
%   \caption{Pion correlator at flow time $t=2t_0$}
%   \label{fig:sub2}
% \end{subfigure}\hfill
% \begin{subfigure}[t]{0.32\textwidth}
%   \centering
%   \includegraphics[width=\linewidth]{corr_flow_22.png}
%   \caption{Pion correlator at flow time $t=5t_0$}
%   \label{fig:sub3}
% \end{subfigure}
% \caption{Pion two-point correlators at different flow times.}
% \label{fig:pion_corr_flow}
% \end{figure}

\section{Summary and outlook}

We have presented the dependence of the pion electromagnetic form factor $F_{\pi}(q^2, t)$ for several values of $q^2$ for different Wilson flow times $t$. We tuned the parameters in such a way that the pion mass was approximately constant along the flow. This is a preliminary step towards a more complete calculation of the $F_{\pi}(q^2, t)$ in the region of parameter space where the QCD vacuum is dominated by instanton contributions. The motivation for this calculation is the possibility of confronting non-perturbative QCD values with the estimation coming from the instanton liquid model.

Several further steps are necessary. Primarily, one should construct the strategy to fix the $c_{SW}$ improvement coefficient along the flow. Although, as we checked explicitly, the value of $c_{SW}$ becomes irrelevant at larger flow times, it considerably affects the pion mass (and other fermionic observables) at $t=0$ and in the region of small $t$. Similarly, if a convenient strategy existed, the vector current improvement coefficient $c_V$ could be determined non-perturbatively along the flow. Eventually, ensembles with other pion masses and lattice spacing should be investigated in order to study the chiral and continuum extrapolations and make contact with the predictions of the instanton liquid model. 

\section*{Acknowledgments}

We thank prof. I. Zahed for inspiration and useful discussions. We also thank S. Zafeiropoulos and K. Cichy for helpful discussions. \\
We gratefully acknowledge Polish high-performance computing infrastructure PLGrid (HPC Center: ACK Cyfronet AGH) for providing computer facilities and support within computational grant no. PLG/2025/017977.\\
We thank Dr. Giovanni Pederiva for providing us with the PACS-CS ensemble of gauge configurations used in this work and for help with the QUDA multigrid solver setup.

\bibliographystyle{unsrt}
\bibliography{references}

%\printbibliography

% \begin{thebibliography}{999}
% \bibitem{zahed} Wei-Yang Liu, Edward Shuryak, and Ismail Zahed, Phys. Rev. D 109,
% 074029 (2024). doi:10.1103/PhysRevD.109.074029

% \end{thebibliography}

\end{document}